\documentstyle[aps,prl,amsfonts]{revtex}
\begin{document}
\twocolumn
\draft
\title{\bf Topology and phase transitions: a paradigmatic evidence}
\author{Roberto Franzosi$^{1,3,}$\cite{roberto}, Marco Pettini$^{2,3,}$
\cite{marco} and Lionel Spinelli$^{4,3,}$\cite{lionel} }

\address{$^1$ Dipartimento di Fisica, Universit\`a di Firenze, Largo E. Fermi 
2, 50125 Firenze, Italy}

\address{$^2$ Osservatorio Astrofisico di Arcetri, Largo E. Fermi 5, 
50125 Firenze, Italy}

\address{$^3$ Istituto Nazionale di Fisica della Materia, Unit\`a di Firenze,
Firenze, Italy} 

\address{$^4$ Centre de Physique Th\'eorique du C.N.R.S., Luminy Case 907,
F-13288 Marseille Cedex 9, France}

\date {\today}
\maketitle
\begin{abstract}
We report upon the numerical computation of the Euler characteristic $\chi$
(a topologic invariant) of the equipotential hypersurfaces $\Sigma_v$ of the
configuration space of the two-dimensional lattice $\varphi^4$ model. 
The pattern $\chi (\Sigma_v)$ {\it vs} $v$ (potential energy) 
reveals that a major topology
change in the family $\{\Sigma_v\}_{v\in\Bbb R}$ is at the origin of the phase
transition in the model considered. The direct evidence given here - of the 
relevance of topology for phase transitions - is obtained through a general 
method that can be applied to any other model.
\end{abstract}
\vskip 1truecm
\narrowtext 
Suitable topology changes of equipotential submanifolds of configuration 
space can entail thermodynamic phase transitions. 
This is the novel result of the present Letter.  The method
we use, though applied here to a particular model, is of general validity and 
it is of prospective interest  to the study of phase transitions in those
systems that challenge the conventional approaches, as it might be the case
of finite systems (like atomic and molecular clusters), of off-lattice 
polymers and proteins, of glasses and in general of amorphous and disordered 
materials. Let us begin by giving a theoretical argument and then proceed 
by numerically proving its truth for the $2d$ lattice $\varphi^4$ model.
Consider classical many particle systems described by standard Hamiltonians 
\begin{equation}
H(p,q)=\sum_{i=1}^N\frac{1}{2}p_i^2 + V(q)
\label{Ham}
\end{equation}
where the $(p, q)\equiv (p_1,\dots ,p_N,q_1,\dots ,q_N)$ 
coordinates assume {\it continuous} \cite{nota0} values and $V(q)$ is bounded
below. The statistical behaviour of physical systems described by Hamiltonians
as in Eq.(\ref{Ham}) is encompassed, in the canonical ensemble, by the 
partition function in phase space 
\begin{eqnarray}
Z_N(\beta )
  &=&\int \prod_{i=1}^N dp_i dq_i e^{-\beta H(p,q)}
 =\left(\frac{\pi}{\beta}\right)^\frac{N}{2}\!\!\int \prod_{i=1}^N dq_i 
e^{-\beta V(q)}   \nonumber \\
&=&\left(\frac{\pi}{\beta}\right)^\frac{N}{2} \int_0^\infty dv\, e^{-\beta v}
\int_{\Sigma_{v}}\frac{d\sigma}{\Vert \nabla V\Vert}\!\!     
\label{zeta}
\end{eqnarray}
where the last term is written using a co-area formula \cite{federer}, and
$v$ labels the equipotential hypersurfaces $\Sigma_v$ of configuration 
space, $\Sigma_v =\{(q_1,\dots,q_N)\in{\Bbb R}^N\vert V(q_1,\dots,q_N)=v\}$.
Equation (\ref{zeta}) shows that for Hamiltonians (\ref{Ham}) the relevant 
statistical information is contained in the canonical configurational 
partition function $Z_N^C=\int\Pi dq_i\exp [-\beta V(q)]$. Remarkably, 
$Z_N^C$ is decomposed -- in the last term of Eq.(\ref{zeta}) -- into an 
infinite summation of geometric integrals,  
$\int_{\Sigma_v} d\sigma\,/\Vert\nabla V\Vert$, defined on the 
$\{ \Sigma_v\}_{v\in{\Bbb R}}$.
Once the microscopic interaction potential $V(q)$ is given, the
configuration space of the system is automatically foliated into the family 
$\{ \Sigma_v\}_{v\in{\Bbb R}}$ of these equipotential hypersurfaces.
Now, from standard statistical mechanical arguments we know that, at any 
given value of the inverse temperature $\beta$, the larger the number $N$ 
of particles the closer to $\Sigma_v\equiv\Sigma_{u_\beta}$ are the 
microstates that significantly contribute to the averages -- computed through 
$Z_N(\beta)$ -- of thermodynamic observables. The hypersurface 
$\Sigma_{u_\beta}$ is the one associated with   
$u_\beta =(Z_N^C)^{-1}\int\prod dq_i V(q) e^{-\beta V(q)}$, the average 
potential energy computed at a given $\beta$.
Thus, at any $\beta$, if $N$ is very large the effective support of the 
canonical measure shrinks very close to a single 
$\Sigma_v=\Sigma_{u_\beta}$.  
Hence, and on the basis of what we found in \cite{cccp,top1,top2}, let us
make explicit the {\sl\underline{Topological Hypothesis}}: 
{\it  the basic origin of a
phase transition lies in a suitable topology change of the $\{ \Sigma_v\}$,
occurring at some $v_c$. This topology change induces the singular behavior
of the thermodynamic observables at a phase transition}. 
By change of topology we mean that  $\{ \Sigma_v\}_{v<v_c}$
are {\it not diffeomorphic} to the $\{ \Sigma_v\}_{v>v_c}$ \cite{diffeo}.
In other words, the claim is that the canonical measure 
should ``feel'' a big and sudden change -- if any -- of the topology  
of the equipotential hypersurfaces of its underlying support, the consequence
being the appearence of the typical signals of a phase transition, i.e.
almost singular (at finite $N$) energy or temperature dependences of 
the averages of appropriate observables.
The larger $N$, the narrower is the effective support of the measure and 
hence the sharper can be the mentioned signals, until true singularities
appear in the $N\rightarrow\infty$ limit.
This point of view has the interesting consequence that -- also at finite 
$N$ -- in principle {\it different} mathematical objects, 
i.e. manifolds of different cohomology type, could be associated to 
{\it different} thermodynamical phases, 
whereas from the point of view of measure theory \cite{LeeYang} 
the only mathematical property available to 
signal the appearence of a phase transition is the loss of analyticity of the 
grand-canonical and canonical averages, a fact which is compatible 
with analytic statistical
measures only in the mathematical $N\rightarrow\infty$ limit. 
In order to prove or disprove the conjectured role of topology, we have 
to explicitly work out adequate information about the topology of the members 
of the family $\{\Sigma_v\}_{v\in\Bbb R}$ for some given physical system.
Below it is shown how this goal is practically achieved by means of numerical
computations. As it is conjectured that the counterpart of a phase transition
is a breaking of diffeomorphicity among the surfaces $\Sigma_v$, it is 
appropriate to choose a {\it diffeomorphism invariant} to probe if and how the
topology of the $\Sigma_v$ changes as a function of $v$. This is a very
challenging task because we have to deal with high dimensional manifolds.
Fortunately a topological invariant exists whose computation is feasible, yet 
demands a big effort. This is the {\it Euler characteristic}, a 
diffeomorphism invariant, expressing fundamental topological information
\cite{pollack}. In order to
make the reader acquainted with it, we remind that a way to analyze a 
geometrical object is to fragment it into other more familiar objects and then
to examine how these pieces fit together. Take for example a surface $\Sigma$
in the euclidean three dimensional space. Slice $\Sigma$ into pieces that
are curved triangles (this is called a triangulation of the surface). Then 
count the number $F$ of faces of the triangles, the number $E$ of edges, 
and the number $V$ of vertices on the tesselated surface. Now, no matter how
we triangulate a compact surface $\Sigma$, $\chi (\Sigma)=F - E + V$ will
always equal a constant which is characteristic of the surface and which is
invariant under diffeomorphisms $\phi :\Sigma\rightarrow\Sigma^\prime$.
This is the Euler characteristic of $\Sigma$. At higher dimensions this can 
be again defined by using higher dimensional generalizations of triangles 
(simplexes) and by defining the Euler characteristic of the $n$-dimensional
manifold $\Sigma$ to be 
\begin{equation}
\chi (\Sigma )=\sum_{k=0}^n (- 1)^k (\#{\rm of~^{\prime\prime} faces~ of~ 
dimension~ k''}). 
\end{equation}
In differential topology a more standard definition of $\chi (\Sigma )$ is
\begin{equation}
\chi (\Sigma )=\sum_{k=0}^n (- 1)^k b_k(\Sigma )   
\label{chi}
\end{equation}
where also the numbers $b_k$ -- the Betti numbers of $\Sigma$ -- are 
diffeomorphism invariants \cite{nota1}. While it would be hopeless to try to 
practically compute $\chi (\Sigma )$ from Eq.(\ref{chi}) in the case of
non-trivial
physical models at large dimension, there is a possibility given by a 
powerful theorem, the Gauss-Bonnet-Hopf theorem, that relates $\chi (\Sigma)$
with the total Gauss-Kronecker curvature of the manifold, i.e. \cite{spivak}
\begin{equation}
\chi (\Sigma)  = \gamma \int_{\Sigma} K_G \,d\sigma
\label{gaussbonnet}
\end{equation}
which is valid for even dimensional hypersurfaces of euclidean spaces
${\Bbb R}^N$ [here ${\rm dim}(\Sigma)=n\equiv N-1$],  and where: 
$\gamma =2/Vol({\Bbb S}^n_1)$ is twice the inverse of 
the volume of an $n$-dimensional sphere of unit radius; $K_G$ is the 
Gauss-Kronecker curvature of the manifold; 
$d\sigma =\sqrt{det(g)}dx^1dx^2\cdots dx^n$  is the invariant volume 
measure of $\Sigma$ and $g$ is the Riemannian metric induced from 
${\Bbb R}^N$.
Let us briefly sketch the meaning and definition of the Gauss-Kronecker
curvature. The study of the way in which an $n$-surface $\Sigma$ curves 
around in ${\Bbb R}^N$ is measured by the way the normal direction 
changes as we move from point to point on the surface. The rate of change
of the normal direction {\boldmath $\xi$} at a point $x\in\Sigma$ in 
direction ${\bf v}$ is described by
the {\it shape operator} $L_x({\bf v}) = - \nabla_{\bf v} 
{\mbox{\boldmath $\xi$}}$, where
${\bf v}$ is a tangent vector at $x$ and $\nabla_{\bf v}$ is the directional
derivative, hence
$L_x({\bf v}) = - (\nabla\xi_1\cdot{\bf v},\dots ,\nabla\xi_{n+1}
\cdot{\bf v})$; gradients and vectors are represented in ${\Bbb R}^N$. 
As $L_x$ is an operator of the tangent space at $x$ into
itself, there are $n$ independent eigenvalues \cite{thorpe} $\kappa_1(x),
\dots,\kappa_n(x)$ which are called the principal curvatures of $\Sigma$ at
$x$. Their product is the Gauss-Kronecker curvature: $K_G(x)=\prod_{i=1}^n
\kappa_i(x)={\rm det}(L_x)$. The practical computation of $K_G$ for the
equipotential hypersurfaces $\Sigma_v$ proceeds as follows. Let
 {\boldmath $\xi$}$=\nabla V/\Vert\nabla V\Vert$ be the unit normal vector to
$\Sigma_v$ at a given point $x$, and let $\{{\bf v}_1,\dots,{\bf v}_n\}$ be 
any basis for the tangent space of $\Sigma_v$ at $x$. Then \cite{thorpe}   
 \begin{equation}
             K_G(x) = \frac{(-1)^n}{\Vert {\nabla V} \Vert^n} 
                \,\left\vert \pmatrix{
                \nabla_{{\bf v}_1} {\nabla V} \cr
                \vdots  \cr
                \nabla_{{\bf v}_n} {\nabla V} \cr
                {\nabla V} \cr }\right\vert\,
                \left\vert
                \pmatrix{
                        {\bf v}_1 \cr
                        \vdots  \cr
                        {\bf v}_n \cr
                        {\nabla V}\cr }\right\vert^{-1}\,.     
  \label{gauss-curv}
        \end{equation}
Let us now consider the family of $\{\Sigma_v\}_{v\in\Bbb R}$ associated with
a particular physical system and show how things  work in practice.
We consider the so-called $\varphi^4$ model on a 
$d$-dimensional lattice ${\Bbb Z}^d$ with $d=1,2$, described by the potential
function 
\begin{equation}
V=\sum_{i\in{\Bbb Z}^d}\left( - \frac{\mu^2}{2}
q_i^2 + \frac{\lambda}{4\!} q_i^4 \right) + 
\sum_{\langle ik\rangle\in{\Bbb 
Z}^d}\frac{1}{2}J (q_i-q_k)^2
\label{potfi4}
\end{equation} 
where $\langle ik\rangle$ stands for nearest-neighbor sites. This system has
a discrete ${\Bbb Z}_2$-symmetry and short-range
interactions; therefore, according to the Mermin-Wagner theorem, 
in $d=1$ there is
no phase transition whereas in $d=2$ there is a symmetry-breaking transition
of the same universality class of the $2d$ Ising model.
Independently of any statistical measure, let us now probe, by computing 
$\chi (\Sigma_v)$ {\it vs} $v$ according to Eq.(\ref{gaussbonnet}),
if and how the topology of the hypersurfaces $\Sigma_v$ varies with $v$. 
To this aim we first devised an algorithm of MonteCarlo type 
by constructing a Markov chain on any desired surface $\Sigma_v$. 
This is obtained by means of a ``demon'' algorithm corrected with a projection
technique \cite{Pettini} which provides a simple and efficient method to 
constrain a random walk on a level-hypersurface, here, of the potential 
function. Each new
step so obtained on $\Sigma_v$ represents a trial step which is accepted or
rejected according to a Metropolis-like ``importance sampling'' criterion
\cite{binder} adapted to the weight $\sqrt{{\rm det}(g)}$.
With any MonteCarlo scheme we can actually compute densities, that is
we can only estimate $\int_{\Sigma_v}K_G d\sigma /\int_{\Sigma_v}d\sigma$, the
average of $K_G$, rather than its total value (\ref{gaussbonnet}) on 
$\Sigma_v$. Hence the need for an estimate of 
$Area(\Sigma_v)=\int_{\Sigma_v}d\sigma$ as a function of $v$. To this aim
we worked out a geometric formula that links the relative
variation of $Area(\Sigma_v)$ with respect to an arbitrary initial value
$Area(\Sigma_{v_0})$, to another MonteCarlo average on $\Sigma_v$: 
$\langle M_1/\Vert\nabla V\Vert\rangle^{\Sigma_v}_{MC}$ where 
$M_1=\frac{1}{n}\sum_{i=1}^n\kappa_i$ is the mean curvature of $\Sigma_v$ 
\cite{nota2}. Thus the final outcomes of our computations are the 
{\it relative} variations of the Euler characteristic.
The computation of $K_G$ at any point $x\in\Sigma_v$ proceeds by working out
an orthogonal basis for the tangent space at $x$, orthogonal to
 ${\mbox{\boldmath$\xi$}}=\nabla V/\Vert\nabla V\Vert$, by means of a 
Gram-Schmidt orthogonalization procedure. Then Eq.(\ref{gauss-curv}) is used 
to compute $K_G$ at $x$. On each $\Sigma_v$ we sampled $1\cdot 10^6\,-\,
3.5\cdot 10^6$ points where we computed $K_G$. This number of points was 
varied,
and several initial conditions were also considered in order to check the 
stability of the results. 
The computations were performed for 
${\rm dim}(\Sigma_v)= 48, 80$ (i.e. $N=7\times 7,\, 9\times 9$) and with the 
choice $\lambda=0.6,\,\mu^2=2,\,J=1$ for the parameters of the potential. 
In order to test the correctness  of our numerical 
``protocol'' to  compute $\chi (\Sigma_v)$, and to assess its degree of
reliability, we 
checked the method against a simplified form of the potential $V$ in Eq.
(\ref{potfi4}), i.e. with $\lambda =J=0,\,\mu^2=-1$. 
In this case the $\Sigma_v$ are
hyperspheres and therefore $\chi ({\Bbb S}^n_v)=2$ for any even $n$.
$Area({\Bbb S}^n_v)$ is analytically known as a function of the radius 
$\sqrt{v}$, therefore the starting value $Area(\Sigma_{v_0})$ is known and 
in this case we can compute the actual values of $\chi (\Sigma_{v})$ instead
of their relative variations only. 
In Fig.1 we report  $\chi (\Sigma_v={\Bbb S}^n_v)$ {\it vs} $v/N$ for 
$N=5\times 5$, the results are in agreement with the theoretical
value within an error of few percents, a very good precision
in view of the large variations of $\chi (\Sigma_v)$ that are found 
with the full expression (\ref{potfi4}) of $V$.
In Fig.2 we report the results for the $1d$ lattice, which is known not to
undergo any phase transition.
Apart from some numerical noise - here enhanced by the more 
complicated topology of the $\Sigma_v$  when $\lambda ,J\neq 0$ -
a monotonously (in the average) decreasing pattern of $\chi (v/N)$ is found. 
Since the variation of $\chi (v/N)$ signals a topology change of the 
$\{\Sigma_v\}$, Fig. 2 tells that a ``smoothly'' varying topology is not
{\it sufficient} for the appearence of a phase transition. In fact, when
the $2d$ lattice is considered, the pattern of $\chi (v/N)$ is very different:
it displays a rather abrupt {\it  change of the topology variation rate with} 
$v/N$ at some $v_c/N$. This result is reported in Fig.3 for a lattice of 
$N=7\times 7$ sites, and in Fig. 4 for a larger lattice of $N=9\times 9$ 
sites \cite{nota3}. The question is now whether the value $v_c/N$, at which 
$\chi (v/N)$ displays a cusp, has anything to do with the thermodynamic phase 
transition, i.e. we wonder if the effective support of
the canonical measure shrinks close to $\Sigma_{v\equiv v_c}$ 
just at $\beta\equiv 1/T_c$, the (inverse) critical temperature of the phase
transition. The answer is in the affirmative. In fact, the numerical 
analysis in Refs.\cite{top1,ccp} shows that -- with $\lambda=0.6,\,\mu^2=2,
\,J=1$ -- 
the function $\frac{1}{N}\langle V\rangle (T)$ and its derivative signal the
phase transition at $\frac{1}{N}\langle V\rangle\simeq 3.75$, a value in very 
good agreement -- within the numerical precision -- with $v_c/N$ where the 
cusp of $\chi (v/N)$ shows up.
Through the computation of the $v$-dependence of a topologic invariant, the 
hypothesis of a deep connection between topology changes  of the 
$\{\Sigma_v\}$ and phase transitions has been given a direct confirmation.
Moreover, we found that a sudden 
``{\it second order variation}'' of the topology of these hypersurfaces 
is the  ``suitable'' topology change - mentioned at the beginning 
of the present Letter - that underlies the phase transition
of second kind in the lattice $\varphi^4$ model. 
There is no reason why the results presented here should be peculiar only 
to the chosen model, and therefore they point to a general validity of the 
relationship between topology and phase transitions, opening a wide field 
of future investigations and applications.
\vskip 0.2truecm

We warmly thank A. Abbondandolo, L.Casetti, C. Chiuderi and G.Vezzosi for 
helpful discussions and comments. One of us (MP) wishes to thank E.G.D. Cohen
and D. Ruelle for an encouraging and helpful discussion held at I.H.E.S.
(Bures-sur-Yvette).

\begin{figure}
\caption{Numerical computation of the Euler characteristic for $24$ 
dimensional spheres. $v$ is the squared radius. }
\label{fig1} 
\end{figure}

\begin{figure}
\caption{1d $\varphi^4$ model. Relative variations of the Euler 
characteristic of $\Sigma_v$ {\it vs} $v/N$ (potential energy density).   
Lattice of $N=1\times 49$ sites. Full line is a guide to the eye.  }
\label{fig2} 
\end{figure}

\begin{figure}
\caption{2d $\varphi^4$ model. Relative variations of the Euler 
characteristic of $\Sigma_v$ {\it vs} $v/N$ (potential energy density).   
Lattice of $N=7\times 7$ sites.
The vertical dotted line corresponds to the phase transition point.
 Full line is a guide to the eye.  }
\label{fig3} 
\end{figure}

\begin{figure}
\caption{2d $\varphi^4$ model. Relative variations of the Euler 
characteristic of $\Sigma_v$ {\it vs} $v/N$ (potential energy density).   
Lattice of $N=9\times 9$ sites.
The vertical dotted line corresponds to the phase transition point.
 Full line is a guide to the eye. }
\label{fig4} 
\end{figure}


\begin{references}
\bibitem{diffeo} A diffeomorphism is a one-to-one differentiable application
with differentiable inverse.
\bibitem{nota2}  $Area(\Sigma_v)=Area(\Sigma_{v_0})
 \exp\int^v_{v_0} du\langle M_1/\Vert\nabla V\Vert\rangle_{MC}^{\Sigma_u}$ 
is the formula we used, it  stems
from $\psi (v)=\log Area(\Sigma_v)$ and the quadrature of
 $\psi^\prime (v)=[d/dv Area(\Sigma_v)]/ Area(\Sigma_v)$ 
written using a geometric derivation formula in Ref.
\protect\cite{federer}.
\end{references}
\end{document}